\documentclass[twocolumn,english,aps,preprintnumbers,amsmath,amssymb,superscriptaddress]{revtex4-2}
\usepackage[latin9]{inputenc}
\usepackage{amsmath}
\usepackage{graphicx}

\makeatletter
\@ifundefined{textcolor}{}
{%
 \definecolor{BLACK}{gray}{0}
 \definecolor{WHITE}{gray}{1}
 \definecolor{RED}{rgb}{1,0,0}
 \definecolor{GREEN}{rgb}{0,1,0}
 \definecolor{BLUE}{rgb}{0,0,1}
 \definecolor{CYAN}{cmyk}{1,0,0,0}
 \definecolor{MAGENTA}{cmyk}{0,1,0,0}
 \definecolor{YELLOW}{cmyk}{0,0,1,0}
}

\def\lsco{La$_{2-x}$Sr$_x$CuO$_4$}
\def\lbco{La$_{2-x}$Ba$_x$CuO$_4$}
\def\lbcoo{La$_{1.885}$Ba$_{0.115}$CuO$_4$}

\usepackage{dcolumn}
\usepackage{bm}
\usepackage{color}

\makeatother

\usepackage{babel}

\begin{document}


\title{Using uniaxial stress to probe the relationship between competing superconducting states in a cuprate with spin-stripe order}


\author{Z.~Guguchia$^{\dag}$}
\email{zurab.guguchia@psi.ch} 
\affiliation{Laboratory for Muon Spin Spectroscopy, Paul Scherrer Institute, CH-5232
Villigen PSI, Switzerland}

\author{D.~Das}
\thanks{These authors contributed equally to the experiments.}
\affiliation{Laboratory for Muon Spin Spectroscopy, Paul Scherrer Institute, CH-5232
Villigen PSI, Switzerland}

\author{C.N.~Wang}
\thanks{These authors contributed equally to the experiments.}
\affiliation{Laboratory for Muon Spin Spectroscopy, Paul Scherrer Institute, CH-5232
Villigen PSI, Switzerland}

\author{T.~Adachi}
\affiliation{Department of Engineering and Applied Sciences, Sophia University, 7-1 Kioi-cho, Chiyoda-ku, Tokyo 102-8554, Japan}

\author{N. Kitajima}
\affiliation{Department of Applied Physics, Tohoku University, 6-6-05 Aoba, Aramaki, Aoba-ku, Sendai 980-8579, Japan}

\author{M.~Elender}
\affiliation{Laboratory for Muon Spin Spectroscopy, Paul Scherrer Institute, CH-5232 Villigen PSI, Switzerland}

\author{F. Br\"{u}ckner}
\affiliation{Institute for Solid State and Materials Physics, Technische Universitat Dresden, D-01069 Dresden, Germany}

\author{S. Ghosh}
\affiliation{Institute for Solid State and Materials Physics, Technische Universitat Dresden, D-01069 Dresden, Germany}

\author{V. Grinenko}
\affiliation{Institute for Solid State and Materials Physics, Technische Universitat Dresden, D-01069 Dresden, Germany}
\affiliation{Leibniz-Institut f\"{u}r Festk\"{o}rper- und Werkstoffforschung (IFW) Dresden, 01171 Dresden, Germany}

\author{T.~Shiroka}
\affiliation{Laboratory for Muon Spin Spectroscopy, Paul Scherrer Institute, CH-5232 Villigen PSI, Switzerland}
\affiliation{Laboratorium f\"{u}r Festk\"{o}rperphysik, ETH Z\"{u}rich, CH-8093 Z\"{u}rich, Switzerland}

\author{M. M\"{u}ller}
\affiliation{Condensed Matter Theory Group, Paul Scherrer Institute, CH-5232 Villigen PSI, Switzerland}

\author{C. Mudry}
\affiliation{Condensed Matter Theory Group, Paul Scherrer Institute, CH-5232 Villigen PSI, Switzerland}
\affiliation{Institute of Physics, \`{E}cole Polytechnique F\`{e}d\`{e}erale de Lausanne (EPFL), CH-1015 Lausanne, Switzerland}

\author{C.~Baines}
\affiliation{Laboratory for Muon Spin Spectroscopy, Paul Scherrer Institute, CH-5232 Villigen PSI, Switzerland}

\author{M.~Bartkowiak}
\affiliation{Laboratory for Scientific Developments and Novel Materials, Paul Scherrer Institut, 5232 Villigen PSI, Switzerland}

\author{Y. Koike}
\affiliation{Department of Applied Physics, Tohoku University, 6-6-05 Aoba, Aramaki, Aoba-ku, Sendai 980-8579, Japan}

\author{A.~Amato}
\affiliation{Laboratory for Muon Spin Spectroscopy, Paul Scherrer Institute, CH-5232 Villigen PSI, Switzerland}

\author{J.M.~Tranquada}
\affiliation{Condensed Matter Physics and Materials Science Division, Brookhaven National Laboratory, Upton, NY 11973, USA}

\author{H.-H. Klauss}
\affiliation{Institute for Solid State and Materials Physics, Technische Universitat Dresden, D-01069 Dresden, Germany}

\author{C.W.~Hicks}
\affiliation{Max Planck Institute for Chemical Physics of Solids, D-01187 Dresden, Germany}

\author{H.~Luetkens}
\email{hubertus.luetkens@psi.ch} 
\affiliation{Laboratory for Muon Spin Spectroscopy, Paul Scherrer Institute, CH-5232 Villigen PSI, Switzerland}

\begin{abstract}

We report muon spin rotation and magnetic susceptibility experiments on in-plane stress effects on the static spin-stripe order and superconductivity in the cuprate system \lbco\ with $x=0.115$. An extremely low uniaxial stress of ${\sim}$ 0.1 GPa induces a substantial decrease in the magnetic volume fraction and a dramatic rise in the onset of 3D superconductivity, from $\sim10$ to 32~K; however, the onset of at-least-2D superconductivity is much less sensitive to stress. These results show not only that large-volume-fraction spin-stripe order is anti-correlated with 3D superconducting (SC) coherence, but also that these states are energetically very finely balanced. Moreover, the onset temperatures of 3D superconductivity and spin-stripe order are very similar in the large stress regime. These results strongly suggest a similar pairing mechanism for spin-stripe order, the spatially-modulated 2D and uniform 3D SC orders, imposing an important constraint on theoretical models.

\end{abstract}


\maketitle

\begin{figure*}[t]
\centering
\includegraphics[width=0.85\linewidth]{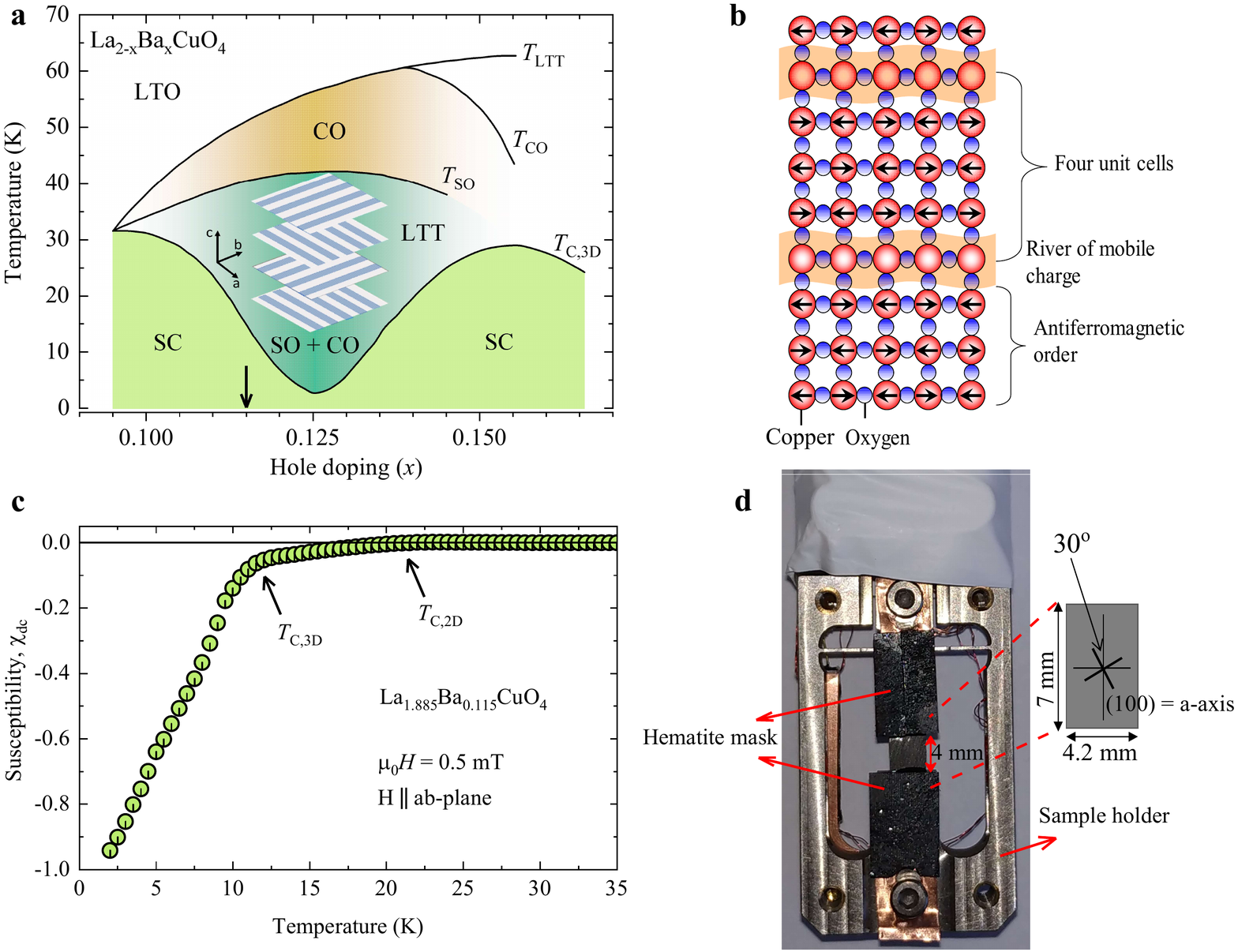}
\vspace{-0.2cm}
\caption{(a) The schematic temperature-doping phase diagram of  La$_{2-x}$Ba$_{x}$CuO$_{4}$.  The arrow indicates the present doping value. The inset illustrates the orthogonal stripe directions between neighbouring layers. The various phases in the phase diagram are denoted as follows: charge-stripe order (CO), low-temperature orthorhombic (LTO), low-temperature tetragonal (LTT), spin-stripe order (SO), and 3D superconductivity (SC). (b) Illustration of a domain of spin- and charge-stripe order for a layer of LBCO, indicating the periods of the charge (4a) and spin (8a) modulations. (c) The temperature dependence of the ZFC magnetic susceptibility for \lbcoo. (d) The uniaxial stress sample holder, used for the ${\mu}$SR experiments.} 
\label{fig1}
\end{figure*}

Cuprate superconductors are believed to exhibit competing superconducting orders: uniform $d$-wave vs. pair density wave (PDW) order \cite{robi19,agte19}. The latter was proposed \cite{berg07} to explain the observation of 2-dimensional (2D) superconductivity with depressed 3D order in \lbco\ (LBCO) near $x=1/8$ with spin-stripe order \cite{li07}.  Whether these states involve distinct electron-pairing mechanisms remains unresolved.  

The conventional BCS theory of superconductivity is based on the Fermi liquid model of electronic states, in which uniformity in real space is assumed and electronic states are characterized entirely by their distribution in reciprocal space. Many discussions of superconducting cuprates have focused only on the nature of the bosonic ``glue'' responsible for electron pairing \cite{aban01,vekh03,dahm09}.  In contrast, others have argued that spatial inhomogeneity is intrinsic to the hole-doped cuprates and a key to understanding the pairing mechanism \cite{emer99,frad15}. Indeed, recent many-body calculations suggest that the uniform and striped (spatially modulated) superconducting states are very close in energy \cite{corb14,zhen17}. At present, the mechanism that controls the competition between such states is still unclear.

Studies of LBCO can provide helpful insight into this unresolved issue, since one of the most astonishing manifestations of competing ordered phases occurs in this system \cite{huck11}. As shown in Fig.~1a, the phase diagram of LBCO exhibits a large dip in the bulk 3D superconducting transition temperature, $T_{\rm c}$, centered at $x=1/8$, coincident with static charge- and spin-stripe order \cite{huck11} (see Fig. 1b). Nevertheless, 2D superconductivity onsets at 40 K, together with spin-stripe order \cite{li07}.  A finite interlayer Josephson coupling would normally be expected to lock the phases of the superconducting wave function between the layers, resulting in 3D order. To explain the apparent frustration of interlayer Josephson coupling, pair-density-wave order within the layers has been proposed \cite{berg07,hime02}, which is compatible with both the charge- and spin-stripe orders.  

What happens when the stripe order is perturbed?  A recent transport study on LBCO $x=0.125$ under strong magnetic fields (applied along the $c$-axis) provided evidence that the putative pairing within the charge stripes is remarkably robust \cite{li19a}. High pressure experiments on LBCO $x=0.125$ have found that the impact on the 3D superconducting transition temperature is quite modest, even beyond the critical pressure where the long-range structural anisotropy, assumed necessary to pin the charge stripes, is absent \cite{huck10,gugu13}.  An optical pump-probe study of LBCO $x=0.115$ found evidence for the suppression of charge-stripe order together with enhanced interlayer superconducting coherence \cite{khan16}; however, the dynamic character of such measurements is not without ambiguity.

Here we perturb a crystal of LBCO $x=0.115$ \cite{Adachi2001} with in-plane compressive stress applied to the CuO$_2$ layers, using an {\it in situ} piezoelectrically-driven stress device \cite{hick18,grin20,Ghosh2020}, while microscopically probing the spin-stripe order with muon spin rotation ($\mu$SR) \cite{Luke1991,Klauss2004,gugu13,sute12,Mohottala,Dalmas1997} spectroscopy and the superconducting transitions with ac susceptibility. The details on the $\mu$SR technique, data analysis, the uniaxial stress device and the sample mounting are given in the supplementary information.

\begin{figure*}[t!]
\centering
\includegraphics[width=0.9\linewidth]{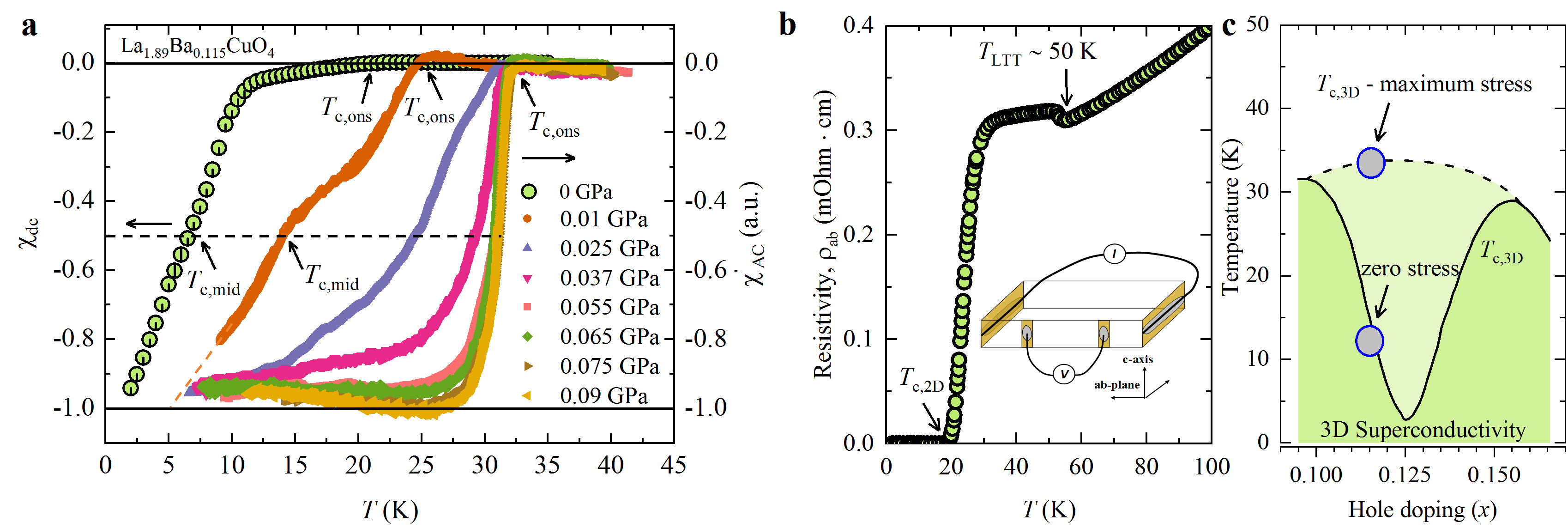}
\vspace{-0.3cm}
\caption{(a) The temperature dependence of the (dia)magnetic susceptibility for \lbcoo, recorded at ambient (left axis) and under various degrees of compressive stress (right axis). 
Arrows mark the onset temperature $T_{\rm c,ons}$ and the temperature $T_{\rm c,mid}$ at which ${\chi}_{\rm dc} = -0.5$. (b) The temperature dependence of the in-plane resistivity (without stress). Electrodes and contacts were placed on the sample as schematically shown in the inset. (c) Schematic temperature-doping phase diagram, indicating the enhancement of 3D SC critical temperature $T_{\rm c,3D}$ under stress for the LBCO $x=0.115$ sample. The value of the $T_{\rm c,3D}$ under maximum stress  is quite similar to the optimal value of SC critical temperature observed in LBCO. The dashed line represents the hypothetical SC phase boundary, expected under applied stress in the broader region around 1/8 doping.} 
\label{fig1}
\end{figure*}

The diamagnetic response of the LBCO $x=0.115$ crystal, measured before mounting in the stress apparatus, is shown in Fig. 1c. The sample was zero-field cooled and then measured in a dc field of $\mu_0 H = 0.5$~mT.
The field was applied parallel to the CuO$_2$ planes, so that the resulting shielding currents must flow between the layers, making the measurement sensitive to the onset of 3D superconductivity near 11 K, consistent with previous work \cite{tran08,gugu16}.  The onset of weak diamagnetism near 22 K corresponds to the 2D superconducting order, as confirmed by the $T$ dependence of the in-plane resistivity (Fig.~2b), which effectively drops to zero at 22~K. Besides the SC transition, an anomaly is seen in the resistivity data at $T_{\rm LTT}$ 50 K (Fig.~2b), which is related to the structural phase transition from a high temperature orthogonal (LTO) to a low temperature tetragonal (LTT) phase.  

A photograph of the $\mu$SR sample holder, which is used to apply uniaxial-stress to the LBCO-0.115 sample, is shown in Fig.~1d. The compressive stress was applied at an angle of 30$^\circ$ to the Cu-O bond direction, denoted as [100].  A previous study of La$_{1.64}$Eu$_{0.2}$Sr$_{0.16}$CuO$_4$ found a rapid enhancement of bulk $T_{\rm c}$ under in-plane uniaxial stress, especially for stress along [110] directions \cite{take04}. To monitor the effect of stress on superconductivity in our case, {\it in situ} ac susceptibility measurements were performed, with an excitation field mostly along the $c$ axis, either just before or after the $\mu$SR measurements, at each stress value.  The results are shown in Fig.~2a.  A comparison with the dc measurement reveals that some stress is present even when the voltage applied to the piezoelectric force generator is zero; possibly due to differential thermal contraction (see Supplementary information for the details of the device). To characterize the changes in superconducting critical temperature, we identify the onset temperature $T_{c,{\rm ons}}$ (which equals $T_{\rm c,2D}$ at zero stress) and midpoint temperature $T_{c,{\rm mid}}$ (which is a good measure of 3D SC order temperature $T_{\rm c,3D}$), as indicated in Fig.~2a , and take the strongest diamagnetic response seen to indicate 100 ${\%}$-volume-fraction superconductivity. As one can see, the compressive stress causes a rapid linear rise of $T_{c,{\rm mid}}$ from 7 to 32~K (with a growth rate of 62.5 K/kbar), where it saturates. The change in $T_{c,{\rm ons}}$ is much more modest. Namely, $T_{c,{\rm ons}}$ increases from 22 to 32~K. Consequently, as indicated in Fig.~2c, the bulk transition $T_{\rm c,3D}$ rises from a very suppressed value to the one that is quite similar to the optimal value of SC critical temperature observed in LBCO or \lsco\ (LSCO) at the same doping level \cite{taka89b}.
  
\begin{figure*}[t!]
\includegraphics[width=0.9\linewidth]{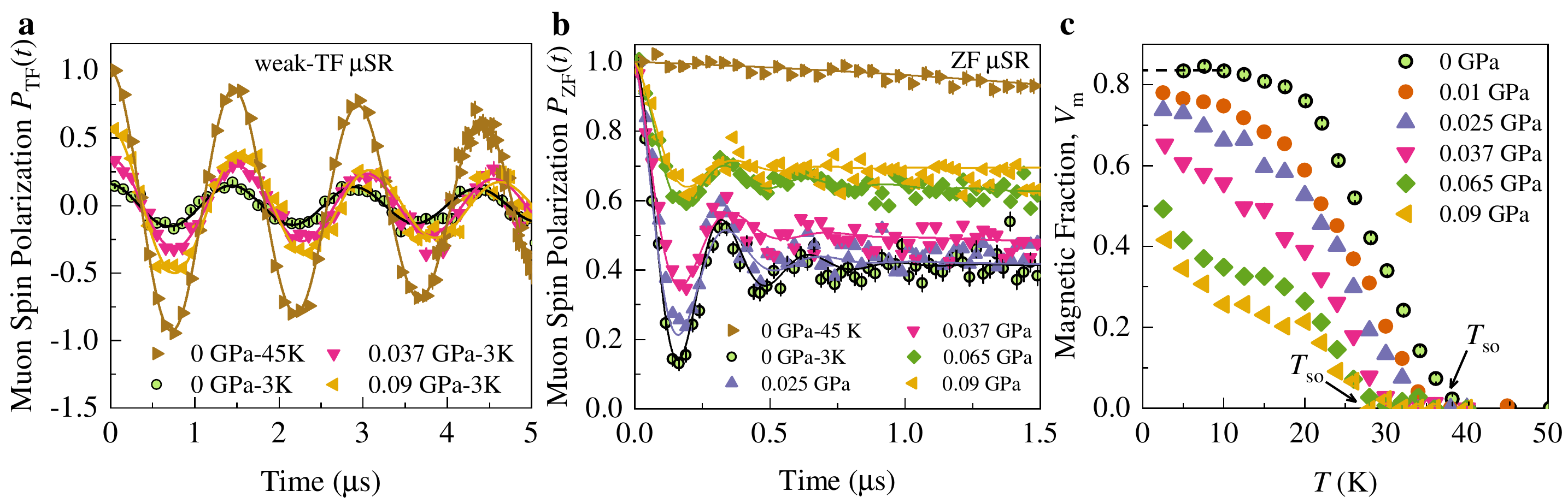}
\vspace{-0.3cm}
\caption{(a) The weak-TF ${\mu}$SR spectra, recorded for \lbcoo at the base temperature $T$ = 3 K under various degrees of compressive stress. (b) The zero-field ${\mu}$SR spectra, recorded at the base temperature under various stresses. (c) The temperature dependence of the magnetically ordered volume fraction recorded under various stresses, as deduced from the TF ${\mu}$SR data shown in panel (a).}   
\label{fig1}
\end{figure*}

The evolution of the spin-stripe order with compressive stress was characterized by a combination of weak transverse-field (TF) and zero-field (ZF) $\mu$SR measurements. In a $\mu$SR experiment, positive muons are implanted into the sample, where each muon spin precesses in the local magnetic field. The time dependent polarization $P(t)$ of the ensemble is monitored by detecting the positrons ejected when the muons decay (see Methods section in supplementary for details). $\mu$SR is an ideal technique for probing materials such as cuprates, where competing phases may exist together and form microscopic inhomogeneity. Measuring the asymmetry between muons counted in detectors on opposite sides of the sample, and then dividing by the maximum possible signal, one obtains the muon polarization function $P_{\rm TF}(t)$, several examples of which are shown in Fig.~3a. In a weak-TF measurement, muons in regions that have no local magnetic order precess in the small applied field. Muons that stop in regions with magnetic order and therefore experience the vector sum of external and internal fields, dephase rapidly. This causes a rapid reduction in the observable $P_{\rm TF}(0)$ (see methods section in Supplementary). Thus, the maximum amplitude of the weak-TF $\mu$SR signal is proportional to the non-magnetic fraction, and the magnetic volume fraction $V_{\rm m}$  can be taken to be $1-P_{\rm TF}(0)$. At 45 K and zero applied stress, $P_{\rm TF}(0) =1$, indicating that there is no magnetic order. At 3 K, $P_{\rm TF}(0)$ is greatly reduced, indicating the development of magnetic order in most of the sample volume.  
Plots of the temperature dependence of $V_{\rm m}$ for various stresses are presented in Fig.~3c. As stress is applied there is a decrease in the spin-ordering temperature $T_{\rm so}$, from ${\sim}$ 38 K at 0 GPa to 
${\sim}$ 30 K at 0.09 GPa. $V_{\rm m}$ decreases much more steeply: at 3 K, by a factor of two and at 10 K, by factor of three at 0.09 GPa.

In ZF $\mu$SR measurements, the muon spins precess exclusively in the internal local field associated with the static magnetic order, with the collective response averaging over the distribution of muon sites relative to the local modulations of the internal field.  As shown in Fig.~3b, several oscillations remain clearly observable under increasing compressive stress values, despite a strong reduction in magnetic volume fraction. The characteristic internal field $B_{\rm int}$ at the muon stopping site can be extracted from the oscillation frequency, as described in the Methods section in the Supplementary information.

Our overall results are summarized in Fig.~4. The spin-stripe order 
temperature $T_{\rm so}$ and superconducting transition temperatures are plotted against stress in Fig.~4a. The stress dependence of the magnetic volume fraction and internal magnetic field at 3 K are shown in Fig.~4b. 
Figure 4a shows that the crossover from 2D to 3D superconducting order occurs at a characteristic uniaxial stress of ${\sigma}_{cr}$ = 0.04 GPa. The dominant change of the spin-stripe order induced by uniaxial stress is a strong reduction in $V_{\rm m}$. $V_{\rm m}$ starts to decrease more rapidly above ${\sigma}_{cr}$, and the reduced $V_{\rm m}$ correlates with the increase (and saturation) of $T_{c,{\rm mid}}$. The 2D-3D crossover has the appearance of a transition that is intrinsically first-order, but broadened by stress inhomogeneity. Further experiments under extremely homogeneous stress conditions are needed to shed light on precise nature of stress induced 2D-3D transition. We note that, only a modest stress-induced decrease in $T_{\rm so}$ (Fig. 4a) and in $B_{\rm int}$ (Fig. 4b) is resolved, indicating that the magnetic structure is well ordered also under stress. Interestingly, $T_{\rm so}$ decreases to essentially match $T_{c,{\rm mid}}$ for ${\sigma}$ ${\textgreater}$ ${\sigma}_{cr}$. There might be a several reasons for the decrease of $B_{\rm int}$ (Fig. 4b): (1) A decrease of the ordered magnetic moment. (2) Slight shift of the muon position due to the modification of the crystal structure. (3) A continuous reorientation of the spin-stripe structure (see supplement), due to a possibly weakened local pinning to the atomic structure as a result of the  applied stress.

To interpret these results, we first recall that the prevalent electronic structure far away from $x=1/8$ is a spatially uniform state, with neither magnetic nor charge order, but with uniform $d$-wave superconductivity. Close to $x=1/8$ a competing phase emerges, with charge and spin stripes pinned along the $a$- and $b$-axes \cite{huck11,axe94}, their orientation alternating from layer to layer \cite{zimm98} (see inset in Fig. 1a). The difference in ordering temperatures for 2D and 3D superconducting order in LBCO with $x$ near $1/8$, as we observe here, implies a strong frustration of the interlayer Josephson coupling. This strong frustration has been rationalized by suggesting PDW order in the layers, with the sign of the superconducting order parameter alternating from stripe to stripe, such that the Josephson coupling between adjacent layers with orthogonal stripes is perfectly geometrically frustrated \cite{berg07,agte19}. Further experimental support for PDW order is provided by recent STM data \cite{ZDU2020}. A perfect stripe phase would, however, suppress the 3D ordering temperature much more than what is observed for LBCO-0.115. This indicates that perfect frustration is probably lifted, either by local deviations from perfect orthogonality of the stripes in adjacent layers, or by the inclusion of patches that remain in the uniform phase. The off-stoichiometric doping in LBCO-0.115 means that local inhomogeneity is likely to be stronger, and patches of uniform superconductivity are likely to be able to establish percolative 3D phase coherence at a higher temperature than at $x$ = 1/8, and indeed, at zero stress $V_{\rm m}$ is 85 ${\%}$, not 100 ${\%}$, showing that the electronic structure of the sample is not homogeneous.

\begin{figure*}[t!]
\centering
\includegraphics[width=0.65\linewidth]{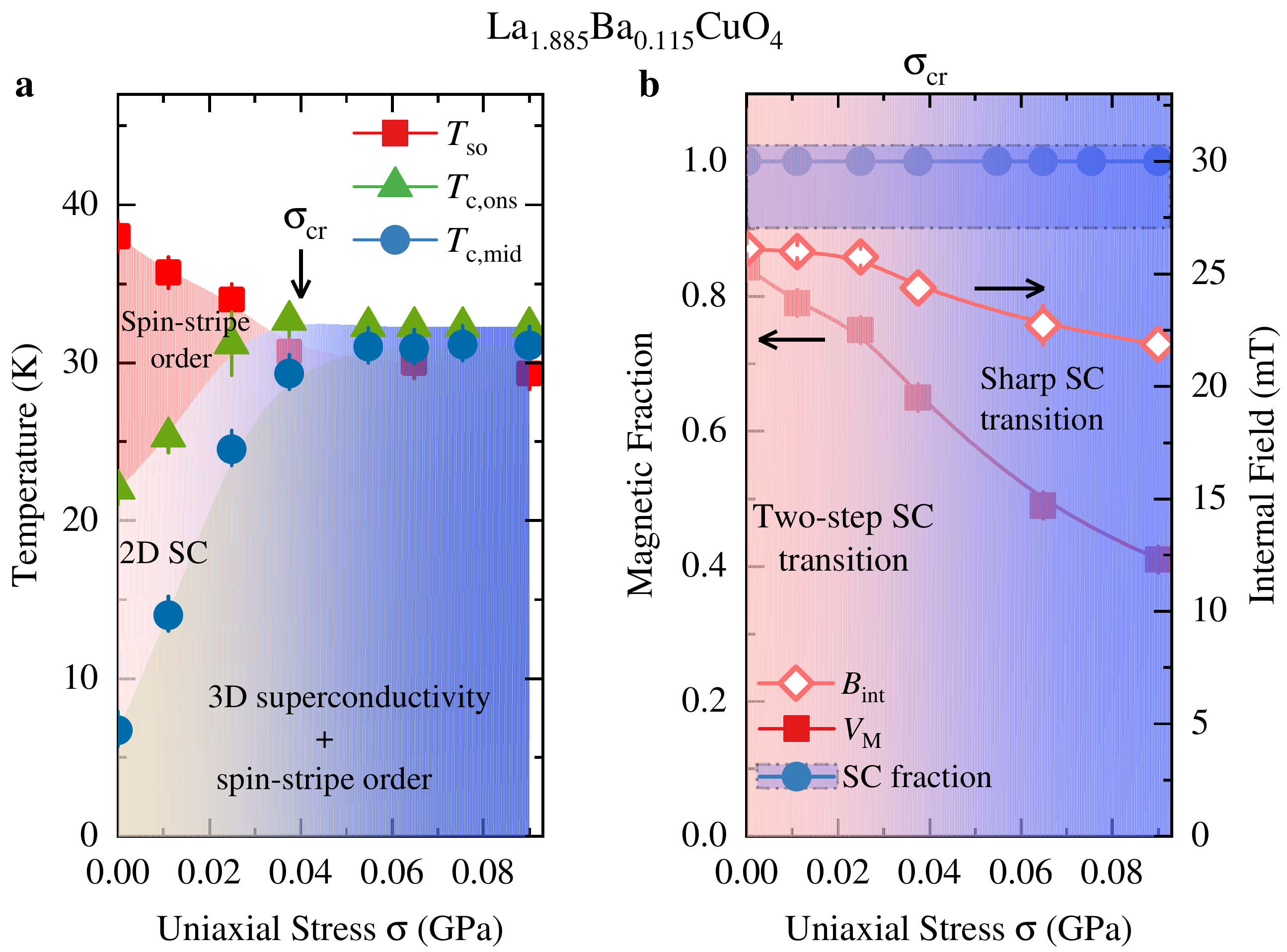}
\vspace{-0.2cm}
\caption{(a) The compressive stress dependence of the SC transition temperatures and of static spin-stripe order temperature $T_{\rm so}$ in LBCO $x = 0.115$. Black arrow marks the critical stress value ${\sigma}_{cr}$, above which a sharp 3D SC transition is established. (b) The stress dependence of the base-$T$ value ($T$ = 3 K) of the magnetically ordered fraction $V_{\rm m}$ and the value of the internal magnetic field $B_{\rm int}$. The SC fraction is only schematic.}
\label{fig1}
\end{figure*}

Applied stress can reinforce both types of deviations from perfect geometric frustration. Since stress distorts the crystal from its tetragonal symmetry, it disfavors orthogonal stripes and thus is expected to promote the abundance of uniform patches. Patches in adjacent layers whose projections overlap, mediate a non-zero interlayer coupling. However, as long as the patches are sparse, the PDW of the stripes dominates the intralayer physics, and the intralayer order parameter has a vanishing uniform-component. Accordingly, the interlayer couplings remain frustrated, very much like in an XY spin glass. These couplings can nevertheless induce an amorphous (glass-like) superconducting 3D order at a finite temperature $T_{c,3D}$, which in general is lower than $T_{c,2D}$. As the fraction of uniform patches increases, $T_{c,3D}$ grows. Beyond a critical fraction of such patches, the superconducting phase will develop a uniform ($Q$ = 0) long-range order both within and between the planes. At that point, $T_{c,3D}$ must coincide with $T_{c,2D}$. 
 
Since spatially-uniform $d$-wave superconducting order in cuprates is empirically known not to show internal static  magnetic order, the scenario of a stress enhanced abundance of uniform patches is consistent with our observation of a  significant decrease in magnetic volume fraction which correlates with the increase of $T_{\rm c,3D}$. A mere reorientation of stripes would instead be hard to reconcile with a decrease in $V_{\rm m}$. Given the drastic change in the superconducting order, it seems likely that the stress reduces the LTT tilting angle \cite{nach98,Klauss2000} or induces a transition to the low-temperature orthorhombic (LTO) phase in some parts of the sample, like the one present in the superconducting phase of LSCO \cite{axe94}, where 3D superconductivity with a similar $T_{\rm c}$ has been observed to coexist with $V_m \approx 20\%$ \cite{savi02}. The observation of nonlinear stress-strain (force-displacement) response (cf. Supplementary Information) provides indirect evidence for structural transitions, that could lead to the formation of additional uniform patches. 
In this context, it is worth pointing out a recent theoretical work on the coexistence of zero and finite momentum superconductivity \cite{Wardh} in which a first order transition between a state with leading PDW order
and sub-leading uniform SC order and a state where the roles are reversed follows naturally in a model with local attraction and repulsive pair hopping.

A key point here is that the variation in onset temperatures of superconductivity, as stress shifts the balance from 2D to 3D superconducting order, is quite modest. This suggests that the underlying (local) pairing mechanisms are essentially the same in the alternative superconducting states with and without spin-stripe order. What evolves instead is the degree to which fluctuations play a role and the way the bulk coherence is established. Remarkably, the stress required to establish the 3D coherence is very small: $\sigma_{cr}$ ${\sim}$ 0.04 GPa (strain of ${\sim}$ 0.05 ${\%}$), which is much smaller than the stress ${\sim}$ 1 GPa (strain of 1 ${\%}$), which is required to, for instance, induce 3D charge density wave order in ${\sim}$ 1/8-doped YBCO \cite{Kim18}. Such tiny stress values are not expected to drive strong changes in the underlying electronic structure in materials such as LBCO. Thus, we conclude that the PDW state in unstressed LBCO-0.115 and the 3D superconductivity in uniaxially stressed LBCO are very close in free energy. Moreover, the onset temperature for the 3D superconductivity and spin-stripe order are quite similar in the not so frustrated large stress regime (beyond the critical stress $\sigma_{cr}$ ${\sim}$ 0.04 GPa), from which we infer that the same kind of electronic interactions are responsible for both phenomena. Given that photoemission studies on LBCO and LSCO at compositions with spin-stripe order indicate the absence of sharply-defined quasiparticle peaks \cite{he09,razz13}, it appears that any realistic theory of the pairing should not rely on Fermi-liquid theory as a starting point.

Our experiment has important implications for the field of high-temperature superconductivity and, hence, should stimulate the development of an adequate theory. It also leads to new questions, such as: What is the impact of the stress on the crystal structure and charge-stripe order? How do these effects vary with doping? How does the transition between PDW and uniform $d$-wave SC states happen? Future experiments will be needed to provide answers. In any case, our results provide a new example of the intriguing behavior that can be uncovered by studies with applied uniaxial stress.

In conclusion, we use muon spin rotation and magnetic susceptibility measurements to follow the evolution of spin-stripe order and superconductivity in LBCO with $x=0.115$ as a function of stress applied within the CuO$_2$ planes. 
We observed that an extremely low uniaxial stress of ${\sim}$ 0.1 GPa causes a substantial reduction of the magnetic volume fraction and a dramatic rise, from $\sim10$ to 32 K, in the onset of 3D superconductivity, while the onset of 2D superconducting order weakly and continuously shifts to the one of the 3D order. Moreover, the onset temperatures for 3D superconductivity and spin stripe order are quite similar in the large stress regime. These results suggest that the underlying pairing mechanisms are essentially the same in the spatially-modulated 2D and the uniform 3D superconducting states, and that the presence of large-volume-fraction spin-stripe order locally inhibits the development of 3D superconductivity.

\textbf{\section{Acknowledgments}}

This work is based on experiments performed at the Swiss Muon Source S${\mu}$S, Paul Scherrer Institute, Villigen, Switzerland. JMT was supported at Brookhaven by the U.S. Department of Energy (DOE), Office of Basic Energy Sciences, Division of Materials Sciences and Engineering, under Contract No.\ DE-SC0012704. This work has been also supported  by the Deutsche Forschungsgemeinschaft (GR 4667/1, GRK 1621, and SFB 1143). T. A. was supported by JSPS KAKENHI through Grant No. JP19H01841. We are grateful to S.A. Kivelson for helpful comments.\\

%

\bibliography{lno,theory,neutrons}

\newpage

\section{Supplementary Information}


\subsection{METHODS}

\subsubsection{Sample preparation} A polycrystalline sample of La$_{2-x}$Ba$_{x}$CuO$_{4}$ with $x$ = 0.115 was prepared by the conventional solid-state reaction method using La$_{2}$O$_{3}$, BaCO$_{3}$, and CuO as precursors. The single-phase character of the samples was checked by powder x-ray diffraction. The single crystal of  La$_{2-x}$Ba$_{x}$CuO$_{4}$ with $x$ = 0.115 was grown by the traveling-solvent floating-zone \cite{Adachi2001}  method. All the measurements were performed on samples from the same batch.\\ 
   
\subsubsection{Principles of the ${\mu}$SR technique} Static spin-stripe order in La$_{2-x}$Ba$_{x}$CuO$_{4}$ with $x$ = 0.115 was studied by means of zero-field (ZF) and weak transverse-field (weak-TF) ${\mu}$SR experiments.
In a ${\mu}$SR experiment an intense beam of 100 ${\%}$ spin-polarized muons is stopped in the sample. The positively charged muons (momentum $p_{\mu}$ = 29 MeV/c) thermalize in the sample at interstitial lattice sites, where they act as magnetic microprobes. In a magnetic material the muon spins precess in the local field $B_{{\rm \mu}}$ at the muon site with a Larmor frequency ${\omega}_{{\rm \mu}}$ = 2${\pi}$ ${\nu}_{{\rm \mu}}$ = $\gamma_{{\rm \mu}}$$B_{{\rm \mu}}$ [muon gyromagnetic ratio $\gamma_{{\rm \mu}}$/(2${\pi}$) = 135.5 MHz T$^{-1}$]. In a ZF ${\mu}$SR experiment, positive muons implanted into a sample serve as an extremely sensitive local probe to detect small internal magnetic fields and ordered magnetic volume fractions in the bulk of magnetic materials. Thus, ${\mu}$SR is a particularly powerful tool to study inhomogeneous magnetism in materials. 

The muons $\mu^{+}$ implanted into the sample decay after a mean life time of ${\tau}$$_{\mu}$ = 2.2 ${\mu}$s, emitting a fast positron $e^{+}$, preferentially along their spin direction. Various detectors placed around the sample track the incoming $\mu^{+}$ and the outgoing $e^{+}$. When the muon detector records the arrival of a $\mu^{+}$ in the specimen, the electronic clock starts. The clock stops when the decay positron $e^{+}$ is registered in one of the $e^{+}$ detectors, and the measured time interval is stored in a histogramming memory. In this way a positron-count versus time histogram is formed. A muon decay event requires that within a certain time interval after a $\mu^{+}$ has stopped in the sample an $e^{+}$ is detected. This time interval extends usually over several muon lifetimes (e.g., 10\,$\mu$s). In the Dolly instrument the sample is surrounded by four positron detectors (with respect to the beam direction): Forward, Backward, Left, and Right. After several millions of muons stopped in the sample, one at a time, one obtains a histrogram for the positrons $e^{+}$ revealed in the forward ($N_{\rm F}$), the backward ($N_{\rm B}$), the left ($N_{\rm L}$) and the right ($N_{\rm R}$) detectors. Ideally, the histrogram counts are described by:

\begin{equation} 
N_{{\alpha}}(t)=N_{0}e^{-\frac{t}{\tau_{\mu}}}[1+A_{0}\vec{P}(t)\vec{n}_{\alpha}]+N_{\mathrm{bg}}. 
\end{equation}
  
 Here, the exponential factor accounts for the radioactive muon decay.
$\vec{P}$($t$) is the muon-spin polarization function with the unit vector 
${\vec{n}_{\alpha}}$ (${\alpha}$ = F, B, L, R) along the direction of the positron detector. $N_{\rm 0}$ is a parameter proportional with the number of the recorded events. $N_{\mathrm{bg}}$ is a background contribution due to uncorrelated starts and stops. $A_{0}$ is the initial decay asymmetry, depending on different experimental factors, such as the detector solid angle, absorption, and scattering of positrons in the material. Typical values of $A_{0}$ are between 0.2 and 0.3. 

  Since the positrons are emitted predominantly in the direction of the muon spin (here precessing with ${\omega_{\mu}}$), the forward and backward detectors will detect a signal oscillating with the same frequency. In order to remove the exponential decay (which reflects simply the muon's finite lifetime), the so-called reduced asymmetry signal $A$(t) is calculated:
\begin{equation} 
A(t)=\frac{N_{F,L}(t)-N_{B,R}(t)}{N_{F,L}(t)+N_{B,R}(t)}=A_{0}P(t),
\end{equation}
where, $N_{F,L}$(t) and $N_{B,R}$(t) are the number of positrons detected in the Forward(Left) and Backward(Right) detectors, respectively. The quantities $A(t)$ and $P(t)$ depend sensitively on the
static spatial distribution and the fluctuations of the magnetic environment of the muons.
As such, these functions reflect the physics of the investigated system \cite{Dalmas1997}.\\

\subsubsection{Analysis of ZF ${\mu}$SR data} The ${\mu}$SR signals (Figure 3b of the main text) in the whole temperature range were analyzed by decomposing the signal into a magnetic and a nonmagnetic contribution: 
\begin{equation}
\begin{split}
P_{ZF}(t)=V_{m}\Bigg[{f_{\alpha} e^{-\lambda_{T}t}J_0(\gamma_{\mu}B_{int}t)}+(1-f_{\alpha})e^{-\lambda_{L}t}\Bigg]  \\
  +(1-V_{m})e^{-\lambda_{nm}t}.
\label{eq1}
\end{split}
\end{equation}
Here, $V_{\rm m}$ denotes the relative volume of the magnetic fraction, and $B_{\rm int}$ is the maximal value of the internal field distribution (Overhauser distribution). ${\lambda}_{\rm T}$ and ${\lambda}_{\rm L}$ are the depolarization rates representing the transverse and the longitudinal relaxing components of the magnetic parts of the sample. $f$ and $(1-f)$ are the fractions of the oscillating and non-oscillating components of the magnetic ${\mu}$SR signal. $J_{0}$ is the zeroth-order Bessel function of the first kind. This is characteristic of an incommensurate spin density wave, as well as of broad internal field distributions with fields ranging from zero to a maximal field and has been regularly observed in cuprates with static spin-stripe order \cite{Luke1991,nach98,gugu13}.
${\lambda_{nm}}$ is the relaxation rate of the nonmagnetic part of the sample, where the spin-stripe order is absent. All the ${\mu}$SR time spectra (both ZF and TF) were analyzed using the free software package {\tt musrfit} \cite{sute12}.\\

\subsubsection{Analysis of weak-TF ${\mu}$SR data} The weak-TF asymmetry spectra, shown in Fig. 3a of the main text, were analyzed by using the function:

\begin{equation}
\begin{aligned}P_{TF}(t)= P_{TF}(0)\exp(-\lambda t)\cos(\omega t + \phi),
\label{eq1}
\end{aligned}
\end{equation}
where $t$ is time after muon implantation, $P_{\rm TF}$($t$) is the time-dependent polarisation, $P_{\rm TF}$(0)
is the initial polarisation (amplitude) of the low frequency oscillating component (related to the paramagnetic volume fraction), ${\lambda}$ is an exponential damping rate due to paramagnetic spin fluctuations
and/or nuclear dipolar moments, ${\omega}$ is the Larmor precession frequency, which is proportional to the strength of the external transverse magnetic field, and ${\phi}$ is a phase offset. As it is standard for the analysis of weak-TF data from magnetic samples the zero for $P_{\rm TF}$($t$) was allowed to vary for each temperature to deal with the asymmetry baseline shift occuring for magnetically ordered samples. From these refinements, the magnetically ordered volume fraction at each temperature $T$ was determined by $V_{\rm m}$ = 1  -- $P_{\rm TF}$(0)($T$). In the paramagnetic phase at high temperature $P_{\rm TF}$(0)($T$ ${\textgreater}$ $T_{\rm so}$) = 1.\\ 

In general, weak-TF signal consists of long-lived oscillations with an applied field and strongly damped oscillations from muons in magnetically ordered regions experiencing a broad field distribution due to the vector sum of applied and internal fields. It is clear from Fig. 3b that in ZF ${\mu}$SR, the depolarisation occurs on a ${\sim}$ 0.5 ${\mu}s$ time scale. However, in TF ${\mu}$SR data, due to a distribution in the angle between the applied and internal fields, the inhomegeneity of the field magnitude is large in the magnetically ordered regions, which causes a strong dephasing of the muons. Therefore and due to a strong data binning which averages the signal in the time bins, the strongly damped signal from muons in magnetic areas of the sample is not visible in weak-TF spectra.\\

\subsubsection{Uniaxial stress device} 

Here, we briefly discuss about the uniaxial stress device and the sample geometry used during the experiments. We used the piezoelectric-driven uniaxial stress device designed for operation at cryogenic temperatures \cite{hick18}, where the sample geometry and sample size are suitable for muon-spin rotation and relaxation experiments. The apparatus fits into the \emph{Oxford~Instruments~Heliox} $^{3}$He cryostat of the general-purpose instrument Dolly on the ${\pi}$E1 beamline at the Paul Scherrer Institute. The sample is mounted in a detachable holder, made of titanium \cite{hick18} (Fig.~5a-b), that allows a rapid sample exchange. The sample holder attaches to the main part of the apparatus, which is called the force generator, and which contains the piezoelectric actuators. The use of piezoelectric actuators allows continuous in situ tunability of the applied pressure. The holder incorporates two pairs of flexures that serve to protect the sample against inadvertent torques and transverse forces. The space around the sample is kept as open as possible, so that muons that do not impact the sample pass through and are picked up by the veto counter (the purpose of  the veto counter is to reject muons and their decay positrons that have missed the sample). The sample plates were masked with hematite (Fe$_{2}$O$_{3}$) foils, which strongly depolarize the incoming stray muons resulting in a loss of the signal. We were able to stop around $\sim40$\%\ of the incoming muons in the sample.\\

 The uniaxial stress device is mainly comprised of two pairs of piezoelectric actuators: one fixed with the device frame (termed as tension stack) and one with the sample (termed as compression stack). The piezoelectric-driven device is preloaded with a force of ${\sim}$ 1000 N using a compression spring. By adjusting the relative voltages in the tension and the compression stacks, we can vary the fraction of the preload applied to the device and hence the sample. The apparatus can only apply compressive forces, whereas, a mechanical gap opens up between the compression stack and the sample holder as soon as the force becomes tensile, constraining the use of our device in tensile stress mode. This is where the sample reaches the zero force position. The displacements and the applied forces were measured using strain gauge bridges (from Tokio Sokki). The instrumental details of our uniaxial stress apparatus are reported in \cite{hick18,grin20}.

We used a La$_{1.885}$Ba$_{0.115}$CuO$_{4}$ sample (referred to as LBCO-0.115 in the following) (7 mm ${\times}$ 4 mm ${\times}$ 0.6 mm) oriented along a crystallographic direction which is off by 30 $^{\circ}$ from
the $a$ axis. The sample was fixed on the sample holder using epoxy (Stycast- 2850 FT). Figure S1 shows the photographs (taken from different viewing angles) of LBCO-0.115 mounted on the sample holder made from Grade-2 Ti. As mentioned in the main article, a pair of coils (each having ~100 turns) was placed very close to the sample for in situ ac-susceptibility (ACS) measurements, which allowed us to determine the $T_{\rm c}$ of LBCO-0.115 under different stress conditions. In order to mask the surface of the sample holder exposed to the muon beam, we used hematite pieces, which strongly depolarizes the incoming muons, hence resulting in loss of asymmetry (signal). The area facing the muon beam was therefore 4 ${\times}$ 4.2 mm$^{2}$. In this way, we were able to stop ${\sim}$ 40 ${\%}$ of the incoming muons in the sample. 

The sample was cooled down while keeping the piezoelectric actuators grounded. We applied the compressive stress at 45 K followed by cooling the system down to 2 K. ACS and ${\mu}$SR measurements were carried out upon warming the sample. In order to apply a compressive stress to the sample, a positive voltage was applied in the compression stack ($V_{\rm C}$). To avoid possible electrical discharges of He gas at 45 K, we had to limit $V_{\rm C}$ to +100 V. To achieve a higher compressive force, we kept $V_{\rm C}$ = +100 V and applied a negative voltage in the tension stack ($V_{\rm T}$). Figure 6 depicts the smooth response of the force sensor as a function of the displacement sensor reading during the application of compressive force to the sample. We observed a small anomaly in the force-displacement curve while $V_{\rm T}$  was decreased from $-25$\,V to $-50$\,V (while keeping $V_{\rm C}$  = 100V). Since we did not observe any signs of damage or slip of the sample, this is likely caused by a change in sensitivity of the lock-in amplifier. We note that the force vs displacement curve is sensitive to a structural transformation. If the stress does not induce a structural transition or a plastic deformation, the force varies linearly with the displacement. In the present case, the force vs displacement is slightly nonlinear at low stresses. The voltage where the linear extrapolations of low- and high-voltage regions cross is in the region between $V_{\rm C}$ = 50--75 V. This may indicate a structural phase transition (full or partial) from a low-temperature tetragonal- to a high-temperature orthorhombic phase around a critical voltage $V_{\rm cr}$ = 50 -- 75 V. The reason for the broad transition around $V_{\rm cr}$ most likely stems from the stress inhomogeneity and from the fact that the set up is not fully optimized to precisely follow the structural phase transition. More on a purely technical level, in some cases the displacement sensor can show strong anomalies as the piezo make contact to the transmission column, i.e., around the zero-force point. Future detailed experiments are needed to explore the structural aspects in detail. 

By using the force-sensor calibration described in detail elsewhere \cite{grin20,Ghosh2020}, we estimated the force applied to LBCO-0.115. We find that 1 ${\mu}$V (change in the force sensor reading from the zero force position) corresponds to ${\sim}$ 30.5 N and the maximum applied force is ${\sim}$ 210 N. Our sample had a cross section area of 2.4 mm$^{2}$, which results in a maximum applied stress of ${\sim}$ 0.09 GPa. Figure 7 shows the calibration curve, where the estimated stress values ${\sigma}$ are plotted as a function of the total applied voltage to the actuators ($V_{\rm C}$ - $V_{\rm T}$).\\

\subsection{Results}
\subsubsection{Uniaxial stress effect on the direction of the internal magnetic field}

 Apart from the magnetic ordering temperature and the magnetic fraction, it is instructive to evaluate the stress dependence of the value, as well as the direction of the internal magnetic field, arising from the spin-stripe order. This was done through zero-field (ZF) ${\mu}$SR experiments. The measurements were performed with the initial muon spin polarisation at an angle of $40^{\circ}$ with respect to the $c$-axis of the crystal, as illustrated in Fig. 8c. The sample was surrounded by two-pairs of detectors: Forward-Backward (F-B) and Left-Right (L-R). Figures~8a and b show representative zero-field (ZF) ${\mu}$SR time spectra for the single crystal LBCO-0.115, recorded using L-R and F-B detectors, respectively. The spectra are recorded at $T$ = 3 K for zero and various applied stresses and at $T$ = 45 K for zero stress. At $T$ = 45 K, the sample is in the paramagnetic state. The paramagnetic state causes only a very weak depolarization of the ${\mu}$SR signal. This weak depolarization and its Gaussian functional form are typical of a paramagnetic material and reflect the occurrence of a small Gaussian Kubo-Toyabe depolarization, originating from the interaction of the muon spin with randomly oriented nuclear magnetic moments. At the base temperature, damped oscillations due to muon-spin precession in internal magnetic fields (with an Overhauser type of field distribution) are observed, indicating the formation of static spin order in the stripe phase of LBCO-0.115 \cite{Luke1991,nach98,gugu13}. In a single crystal, the amplitude of the oscillation is proportional to two quantities: the magnetically ordered volume fraction $V_{\rm m}$ and the angle between the muon spin polarisation and the internal field ${\theta}$ (see Fig. 8d, where the muon spin precession around the internal magnetic field, for two extreme cases $B_{\rm int}$ ${\perp}$ $c$ and $B_{\rm int}$ ${\parallel}$ $c$, is schematically illustrated). For the internal field direction, shown in the top panel of Fig. 8d, the ${\mu}$SR signal from F-B detectors exhibits the maximum amplitude and no oscillations will be detected in the L-R detectors. The opposite will be observed for the configuration shown in the bottom panel of Fig. 8d. Thus, by evaluating the data from all four detectors one can obtain useful information on the direction of the internal field and, therefore, the local spin configuration. Figs. 8a-b show that the application of stress causes a substantial reduction of the signal amplitude. A careful inspection of the data from all four detectors reveals that the strong effect on the amplitude cannot be fully explained by the reduction of the volume fraction, as evidenced by weak-TF ${\mu}$SR, shown in the main text. In part, it results from a change of the direction of the internal field at the muon stopping site. Taking into account the independently determined volume fraction (from weak-TF ${\mu}$SR) for the analysis of the zero-field ${\mu}$SR data, we can extract the fraction of the oscillating components $f_{\rm LR}$ and $f_{\rm FB}$, for L-R and F-B detectors, respectively. Both such components depend only on the angle ${\theta}$ between the internal field and the muon spin polarisation. These quantities should be constant with respect to the applied stress, as long as the direction of the internal field does not change. Thus, they provide qualitative information about the internal field orientation and therefore, the local spin structure. As shown in Fig. 9, $f_{\rm LR}$ and $f_{\rm FB}$ show a non-monotonous change with stress, implying a change in the direction of the internal field with stress. This, in turn, evidences that with stress the stripe direction is somewhat modified. At present, we do not have a quantitative understanding of how stress modifies the stripe direction, which opens the way to rigorous experimental and theoretical studies to fully exploit this aspect. We note also that a change of the muon position due to structural changes might contribute to the observed effect.

\begin{figure*}[t!]
\includegraphics[width=0.75\linewidth]{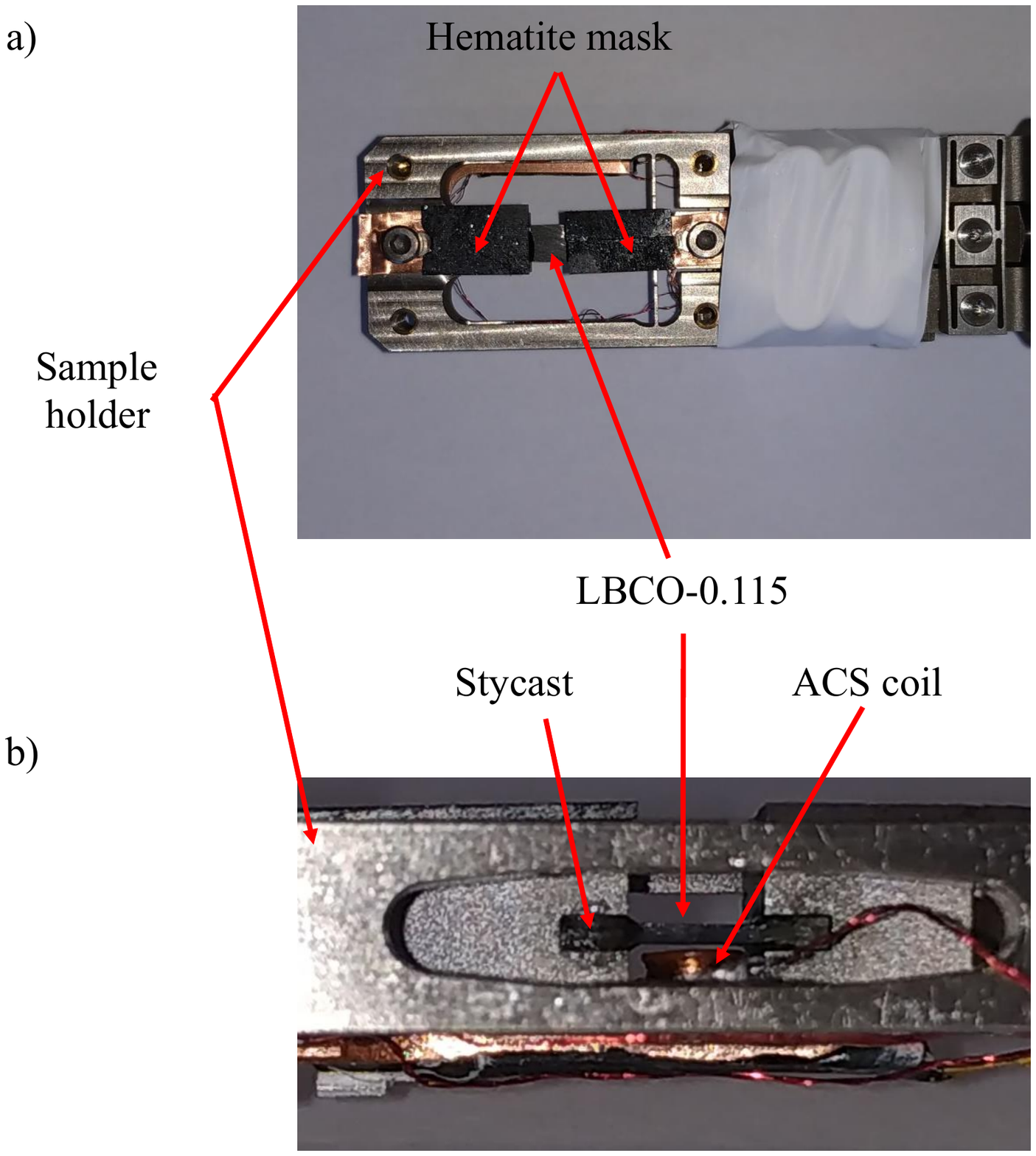}
\vspace{-4.0cm}
\caption{Photograph of the LBCO-0.115 crystal mounted on the sample holder. (a) View from the direction of the incoming muon beam. Hematite pieces masking the holder frame exposed to the muon beam. (b) Side view of the sample showing ACS coil.}
\label{fig1}
\end{figure*}

\begin{figure}[t!]
\includegraphics[width=1.0\linewidth]{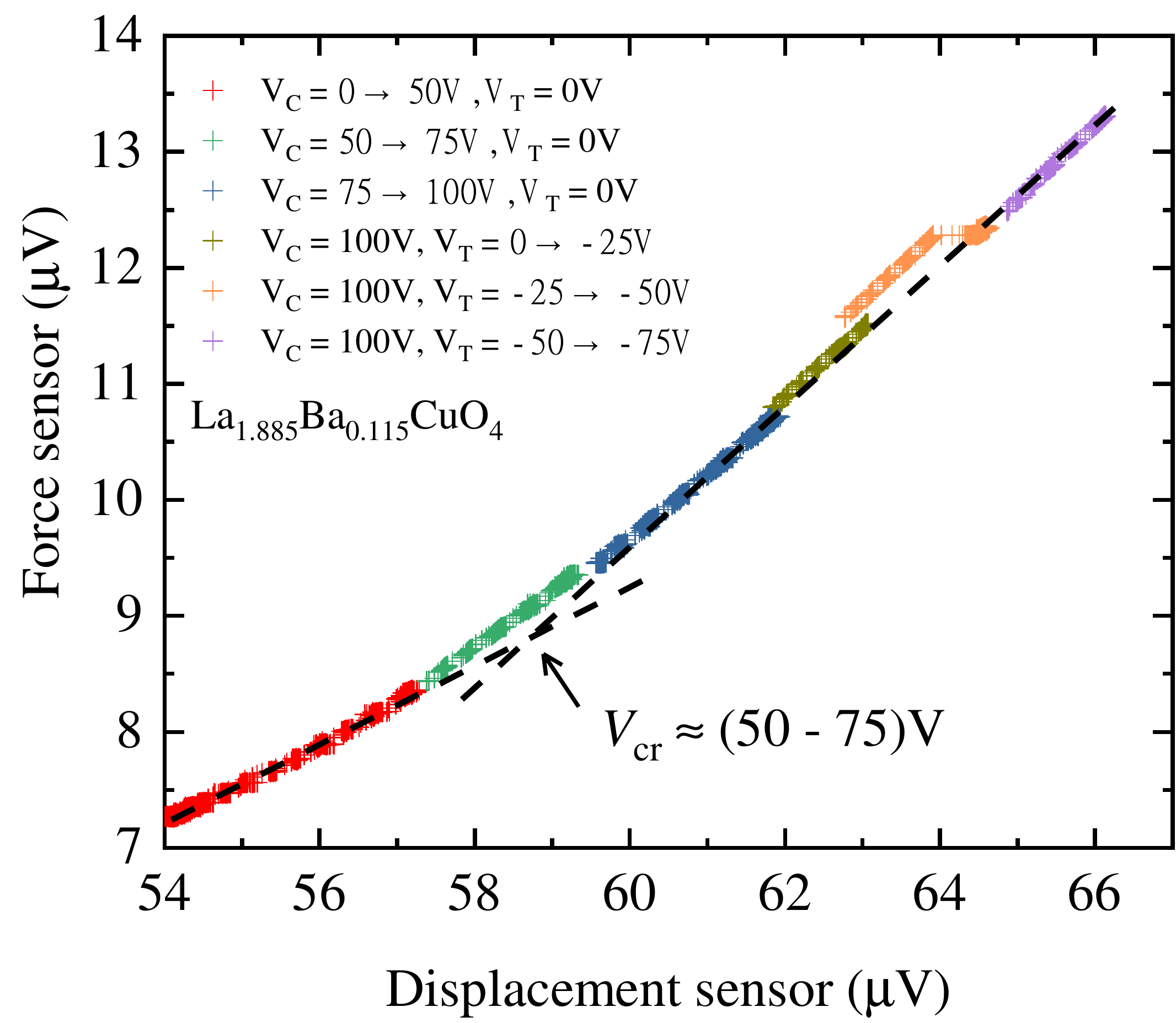}
\caption{Force sensor reading as a function of displacement sensor reading when a compressive force is applied. The dashed lines are linear fits to the data in the low and high voltage regions.}
\label{fig1}
\end{figure}

\begin{figure}[t!]
\includegraphics[width=1.0\linewidth]{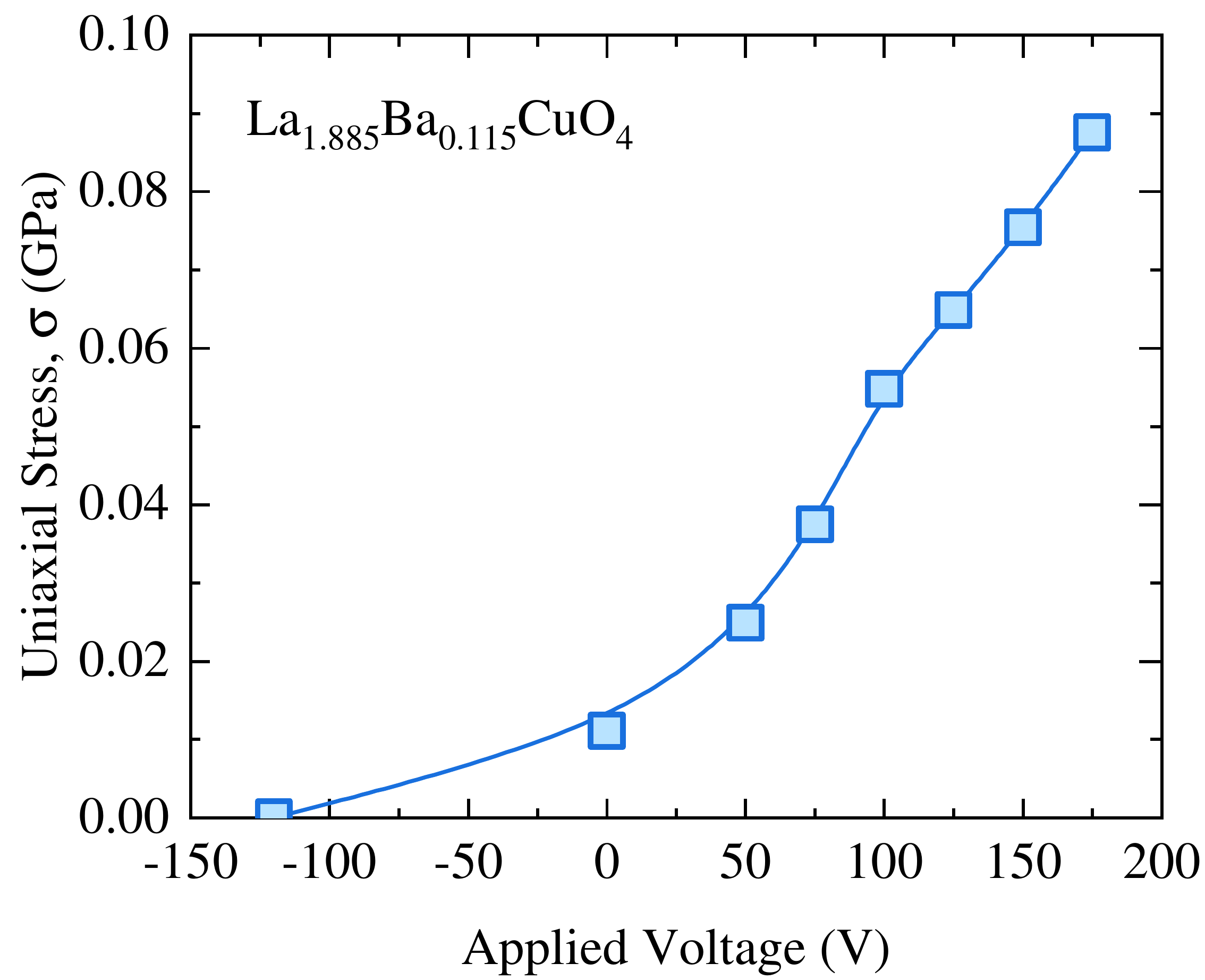}
\caption{Estimated stress values as a function of the total applied voltage to the actuators.}
\label{fig1}
\end{figure}

\begin{figure*}[t!]
\includegraphics[width=0.8\linewidth]{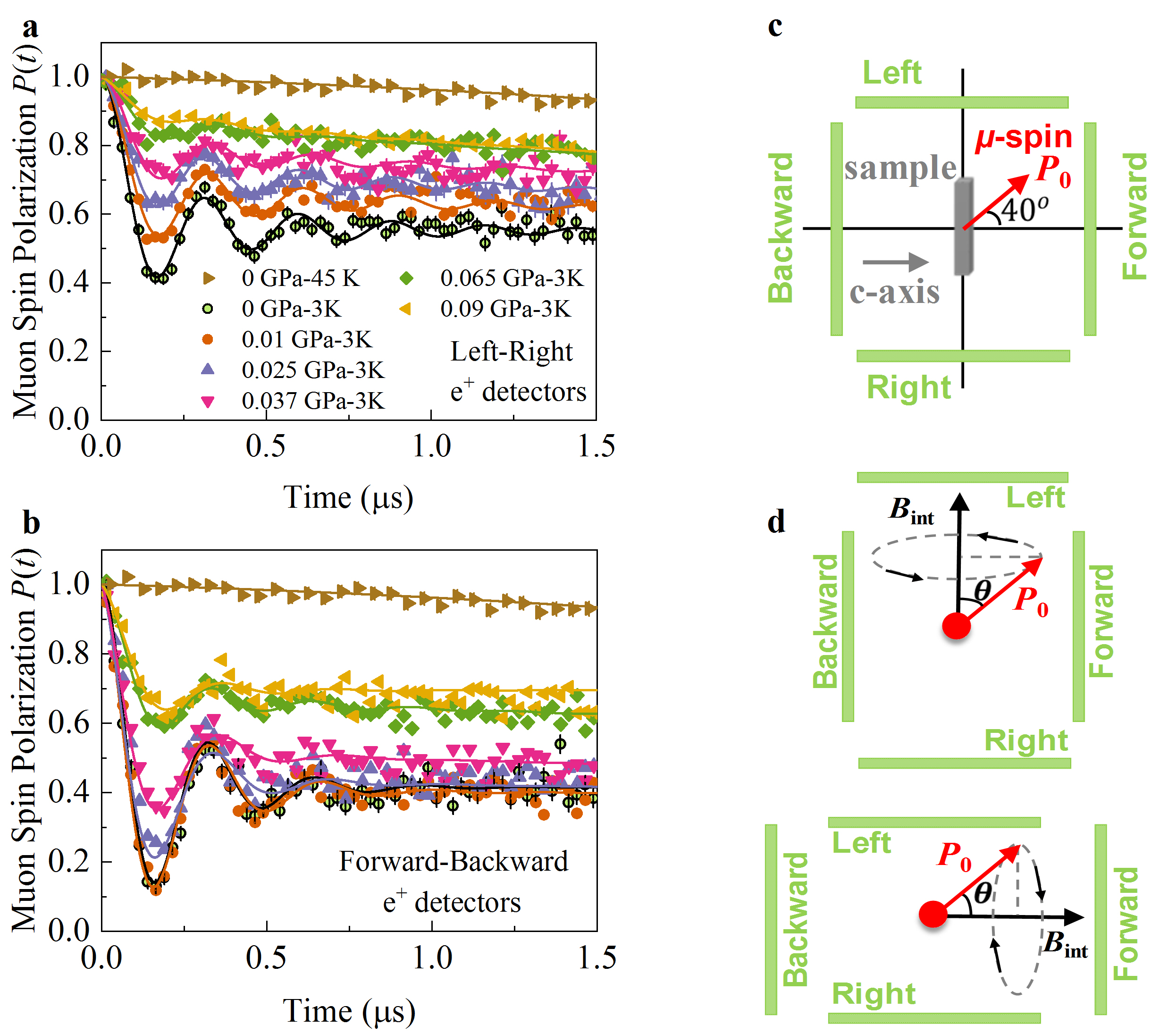}
\caption{(a,b) The zero-field ${\mu}$SR spectra, recorded at the base temperature under various stress values. (c) A schematic overview of the experimental setup for the muon spin forming $40^{\circ}$ with respect to the $c$-axis of the crystal. The sample was surrounded by four detectors: Forward (F), Backward (B), Left (L) and Right (R). (d) Schematic illustration of the muon spin precession around the internal magnetic field for two cases: (top) The field is perpendicular to the $c$-axis and points towards the L-detector. ${\theta}$ is the angle between the magnetic field and the muon spin polarization at $t$ = 0. (bottom) The field is parallel to the $c$-axis of the crystal and points towards the F-detector.}
\label{fig1}
\end{figure*}

\begin{figure}[t!]
\includegraphics[width=1.0\linewidth]{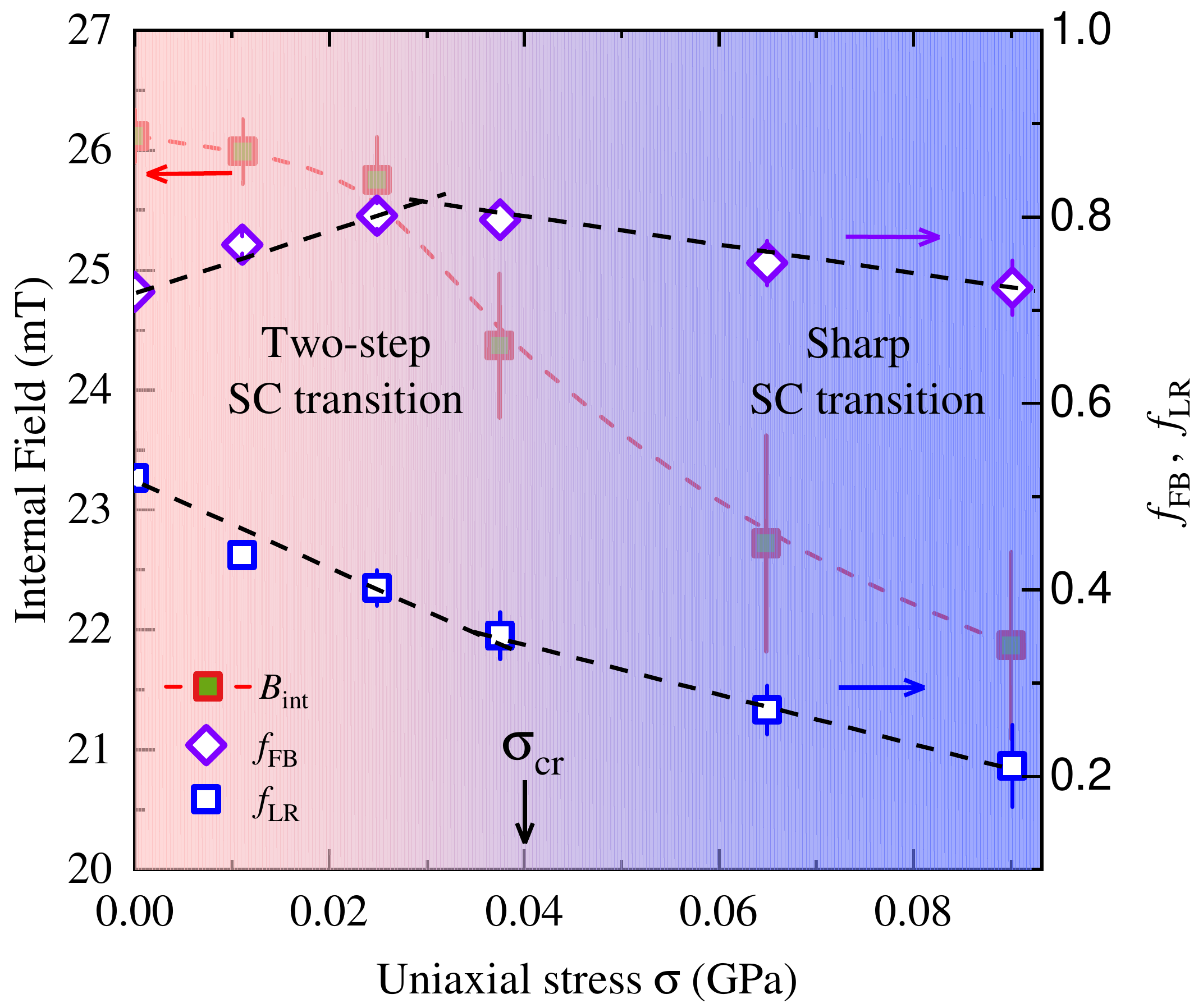}
\caption{The stress dependence of the internal magnetic field value and the ${\theta}$ dependent part of the oscillating ${\mu}$SR signal fractions, $f_{FB}$ and $f_{LR}$, for F-B and L-R detectors, respectively. The quantities $f_{FB}$ and $f_{LR}$ depend on the direction of the internal field.}
\label{fig1}
\end{figure}

\end{document}